\documentclass[twocolumn,prl,10pt,superscriptaddress]{revtex4-1}
\usepackage[latin1]{inputenc}
\usepackage[dvips]{graphicx}

\usepackage{dcolumn}
\usepackage{bm}
\usepackage{dsfont}
\usepackage{color}
\usepackage{url}
\usepackage[colorlinks=true ,citecolor=blue]{hyperref}
\usepackage{amsmath,amssymb,esint}
 \usepackage[table]{xcolor}

\definecolor{Nathanblue}{rgb}{0.,0.24,0.51}

\newcommand{\be}{\begin{equation}}
	\newcommand{\ee}{\end{equation}}
\newcommand{\bq}{\begin{eqnarray}}
	\newcommand{\eq}{\end{eqnarray}}

\newcommand{\tr}{\mathrm{tr}}

\newcommand\beq{\begin{equation}}
\newcommand\eeq{\end{equation}}
\newcommand\bal{\begin{aligned}}
\newcommand\eal{\end{aligned}}

\begin{document}

\title{Fractional Quantum Hall Effect for Extended Objects:\\ from Skyrmionic Membranes to Dyonic Strings}

\author{Giandomenico Palumbo}
\affiliation{School of Theoretical Physics, Dublin Institute for Advanced Studies, 10 Burlington Road,
	Dublin 4, Ireland}

\date{\today}

\begin{abstract}
\noindent It is well known that in two spatial dimensions the fractional quantum Hall effect (FQHE) deals with point-like anyons that carry fractional electric charge and statistics. Moreover, in presence of a SO(3) order parameter, point-like skyrmions emerge and play a central role in the corresponding quantum Hall ferromagnetic phase.
 In this work, we show that in six spatial dimensions, the FQHE for extended objects shares very similar features with its two-dimensional counterpart. In the higher-dimensional case, the electromagnetic and hydrodynamical one-form gauge fields are replaced by three-form gauge fields and 
the usual point-like anyons are replaced by membranes, namely two-dimensional extended objects that can carry fractional charge and statistics. We focus on skyrmionic membranes, which are associated to a SO(5) order parameter and give rise to an higher-dimensional generalizaton of the quantum Hall ferromagnetism. We show that skyrmionic membranes naturally couple to the curved background through a generalized Wen-Zee term and can give us some insights about the chiral conformal field theory on the boundary. We then present a generalization of the Witten effect in six spatial dimensions by showing that one-dimensional extended monopoles (magnetic strings) in the bulk of the FQH states can acquire electric charge through an axion field by becoming dyonic strings.
\end{abstract}

\maketitle
\section{Introduction}
\noindent Higher-dimensional topological matter represents a very active field of research due to the possibility to engineer synthetic dimensions in suitable artificial setups \cite{Celi, Ozawa,Yuan, Prodan, Kraus, Ezawa}.
In fact, topological invariants in four and higher dimensions and novel quantum effects such as the higher-dimensional Thouless pumping and anomalous quantum transports can be experimentally detected and measured in synthetic matter \cite{Zhang, Goldman, Carusotto, Bloch, Zilberberg, Petrides, Lee, Sacha, Zhu, Palumbo, Yu, Chen, Weisbrich}.
At microscopic level, these phases deal with point-like quasiparticles. In fact, in the higher-dimensional integer QH states, the corresponding topological quantum field theories are built from one-form gauge fields and describe the physical features of quasi-particles in the low-energy regime \cite{Karabali1,Karabali2,Karabali3,Karabali4, Karabali5, Tong}. This picture can be also extended in the interacting regime \cite{Zhang}. In this work, we highlight a complementary point of view, in which higher-dimensional phases represent the natural playground where extended objects naturally emerge.
For instance, when one-form gauge fields are replaced by three-form fields, a 6D FQHE for membranes can be consistently built \cite{Wu-Zee, Tze, Bernevig, Hasebe, Heckman}. 
These emergent membranes, which naturally couple to three-form gauge fields, could in principle emerge from suitable higher-dimensional lattice microscopic models similarly to the string-like and membrane-like vortices that naturally emerge in higher-dimensional superfluids and superconductors \cite{Price, Sarkar, Pretko}. Moreover, besides topological phases, a generalization of Anderson localization for strings and membranes has been recently formulated \cite{Pretko}.
It is then worth to further investigate the physics of extended objects, such as membranes and strings and figure out all their physical implications in higher-dimensional topological matter.\\

\noindent The main goal of this work is to unveil novel physical effects related to membrane and string configurations in the 6D FQHE. We will first remark all the common features between the 6D FQHE for anyonic membranes and the 2D FQHE for point-like anyons by discussing the corresponding Hall conductivity, fractional statistics, bulk-edge correspondence and geometric features in curved space.
We will start considering the topological-quantum-field-theory description of the FQHE in terms of tensorial Chern-Simons theories and the generalized anomaly inflow.
We will then show that there exist suitable SO(5) skyrmionic configurations that give rise to a 6D quantum Hall ferromagnetism in analogy to the two-dimensional case. 
Moreover, these skyrmionic membranes will give us some insights about the chiral 5+1-D CFT that lives on the boundary on the system and carries a 't Hooft (gravitational) anomaly.
We will then consider string-like objects (extended monopoles) carrying magnetic charge and show that they acquire an electric charge by becoming dyons through a generalized Witten effect. All these points are summarized in Table \ref{Table_one}, where we have also included the 4D case, which is very different with respect to the 2D and 6D cases. In fact, the 4D FQHE for strings is not well defined for a number of reasons that will be discussed in the paper. In the Appendix A, we will provide a more detailed discussion about the 4D case by showing the existence of dyonic strings (this is one of the few features shared with the 6D case).\\
Although, our results are based on effective quantum field theories in the continuum, the corresponding physical implications rely on observables that could be in principle measured in suitable synthetic systems described by these effective field theories in the low-energy limit.

\section{6D FQHE for membranes}
We start revisiting and summarizing the main features of the 6D FQHE for membranes by focusing on its topological and geometric responses. 
The effective topological quantum field theory for the Laughlin states of membranes has been introduced in Ref.\cite{Heckman}
\beq\label{CSaction2}
S_{\rm {CS}}[c, C]=\int_{M_7} \left[-\frac{p}{4\pi} c \wedge dc +\frac{1}{2\pi} C \wedge dc+ J \wedge \star C\right],
\eeq
where $p$ is an odd-integer number, $\star$ is the Hodge symbol (it contains the information of the volume element in curved space), $J$ is a three-form current associated to membranes, $C$ an external three-form gauge field, while $c$ is a three-form hydrodynamic field.
These tensorial Chern-Simons theories and their quantization have been already discussed in the high-energy-physics literature from different points of view \cite{McGreevy, Gukov1, Gukov2, Krasnov}. 
Similarly to the 2D FQHE, we can integrate out the emergent gauge field, by obtaining
\beq\label{CSaction3}
S_{\rm {CS}}[C]=\int_{M_7} \left[\frac{1}{4\pi p} C \wedge dC+J \wedge\star C\right],
\eeq
which gives us the covariant Hall current
\beq\label{Hall}
J = \frac{1}{2\pi p} \star dC.
\eeq
Here, $D=dC$ is the four-form field strength and in components, $B^{ij}=\epsilon^{ijklmn} D_{klmn}$ and $E_{ijk} = D_{ijk0}$ play the role the (tensorial) magnetic and electric fields, respectively.
In this picture, the membranes are electrically charged with respect to $E$ and fractional conductivity for membranes is manifest in the above expression. 
The anyonic nature of these membranes and their corresponding fractional charge can be rigorously established by analyzing their fractional statistics through a generalized (tensorial) flux attachment and Gauss-Hopf linking number calculated by considering the membrane world volumes \cite{Tze,Hasebe,Bernevig}.
Here, the world volumes of the anyonic membranes naturally replace the world lines of point-like anyons of the lower-dimensional case.\\
When $M_7$ has a spatial boundary $M_6=\partial M_7$ then the boundary states of the 6D FQH states are described by a 5+1-D chiral conformal field theory (CFT) characterized by a (anti-)chiral three-form field strength, namely $\mathcal{H}=\pm \star \mathcal {H}$, where $\mathcal{H}=d\mathcal{B}$ and $\mathcal{B}$ the two-form gauge potential \cite{Heckman,McGreevy, Nair}. This theory naturally generalizes the 1+1-D chiral boson that lives on the boundary of the 2D Laughlin states.
This generalized bulk-boundary correspondence can be understood through the anomaly inflow of chiral higher-form gauge theories as recently discussed in Ref.\cite{Tachikawa}.
We point out that this bulk-edge correspondence though higher-forms and anomaly inflow does not apply to an hypothetical 4D FQHE for extended objects (strings), in which the 3+1-D boundary theory does not possess any gravitational anomaly differently from the 2D and 6D cases \cite{Witten}. Moreover, differently from Eq.(\ref{CSaction3}), a tensorial Chern-Simons theory for a single two-form field $\mathcal{B}$ in 4+1-D dimensions is a total derivative
and does not provide any (fractional) Hall current in the bulk \cite{McGreevy}.
In 4+1-D, only time-reversal-invariant BF-like theories are still well defined in the bulk and give rise to 3+1-D (non-chiral) dynamical gauge theories on the boundary \cite{McGreevy,Magnoli}. Thus, the 4D FQHE can have a physical realization only in terms of point-like quasiparticles and the corresponding effective topological action is given in terms of Chern-Simons terms built from one-form gauge fields \cite{Karabali1,Karabali2,Karabali3}.
We now show further similarities between the 2D and 6D FQHEs by introducing a curved-space background. 
In the 2D FQHE case, it is well known that there exists a non-trivial coupling between the Abelian SO(2) spin connection $\omega$ and the electromagnetic field $A_{em}$, i.e. $\omega \wedge d A_{em}$. This is known as Wen-Zee term and plays a central role in the study of Hall viscosity \cite{Wen-Zee, Cappelli}. In the 6D case, we propose the following generalized Wen-Zee-like term
\beq\label{CSaction4}
S_{\rm {WZ}}[C, \omega]=\frac{1}{4\pi p}\int_{M_7} {\rm \tr}\, {\rm CS}_3(\omega) \wedge dC,
\eeq
where ${\rm CS}_3(\omega)=\omega \wedge d\omega + (2/3) \,\omega \wedge \omega \wedge \omega$, with $\omega$ here the spatial $SO(6)$ spin connection, which is intrinsically non-Abelian.
The above expression is obtaining after integrating out the hydrodynamic field $c$ that directly couples with the spin connection in the FQH regime through a similar mixed topological term. Notice that due to $C$ field, the above action is different from the generalized Wen-Zee term derived in Ref.\cite{Karabali4} for the 6D QHE of point-like quasi-particles. When the spin connection and frame field $e$ are independent then the torsion tensor $T= D e + \omega \wedge e$ is null zero anymore and a further mixed topological term can be built
\beq\label{CSaction5}
S_{\rm {T}}[C, \omega, e]=\frac{\beta}{4\pi p}\int_{M_7} {\rm \tr}\, {\rm CS}_3(e) \wedge dC,
\eeq
where $\beta$ is a dimensional parameter and ${\rm CS}_3(e)=e \wedge T$. Notice, that both the gravitational ${\rm CS}_3(\omega)$ and torsional Chern-Simons terms ${\rm CS}_3(e)$ have been already appeared in the context of topological insulators and superconductors in lower dimension \cite{Fradkin, Hughes, Palumbo6, Nissinen, Stone} and their linear combination in 2+1-D represents the action for exotic gravity \cite{Witten2}, where $\beta$ plays the role of cosmological constant. Thus, the 2+1-D exotic gravitational action could be induced from the above terms by employing a suitable dimensional compactification. 
Due to the two above terms, the covariant Hall current in Eq. (\ref{Hall}) is modified as follows
\begin{align}
J = \frac{1}{2\pi p} \star dC + \frac{1}{4\pi p} \star {\rm \tr} \left(R \wedge R\right)+\nonumber \\ \frac{\beta}{4\pi p} \star {\rm \tr} \left(T \wedge T- R \wedge e \wedge e\right),
\end{align}
where $R=D_\omega \omega= d \omega+ \omega \wedge \omega $ is the two-form Riemann tensor.
This is a perfectly gauge-invariant expression, however,
differently from the 2D case \cite{Wen-Zee}, the above geometric terms do not contribute through an Euler number associated to the six-dimensional space once the electric density is integrated on the whole space.
However, when $\star J$ is integrated on a four-dimensional compact manifold, then the two geometric terms give us the first Pontryagin invariant \cite{Zanelli}.
Importantly, purely geometric contributions such as those ones considered in Refs \cite{Karabali4,Karabali5} would need to be included in the effective topological action in order to also take into account the gravitational anomaly of the boundary.\\
Finally, we want to remark that besides topological and geometric responses and the bulk-boundary correspondence, the FQHE is an incompressible quantum fluid. This central bulk feature in 2D is encoded the $W_{\infty}$ or GMP algebra \cite{GMP,Cappelli2,Karabali} , which is related the underlying quantum area-preserving diffeomorphisms induced by interactions. This can be also understood by employing noncommutative geometry. As recently shown in Ref.\cite{Hasebe}, noncommutative geometry also appears in the higher-dimensional QHE and is deeply related to Nambu geometry. In other words, membranes and $C$ fields give rise to a noncommutative geometry and this would represent a signature of an emergent volume-preserving diffeomorphisms for membranes.
This represents a further evidence that the 6D FQHE shares (almost) all the main features of the 2D counterpart.
In the next sections we will also discuss some novel features (skyrmionic configurations) that are instead unique for the 6D case.

\begin{table}[!ht]
	\renewcommand{\arraystretch}{2}
	\begin{tabular}{|c|c|c|c|}
		\hline
		FQHE for $(D/2-1)$--dimensional objects    & 2D & 4D & 6D  \\ \hline
		Chern-Simons terms for $D/2$ forms           & $\bigcirc$                              & $\times$                         &                $\bigcirc$                               \\ \hline
		Mixed Chern-Simons terms & $\bigcirc$                                & $\bigcirc$                             & $\bigcirc$          \\ \hline
		Hopf map        & $\bigcirc$                             &  $\times$                             & $\bigcirc$                                               \\ \hline
		Wen-Zee term for $D/2$ forms & $\bigcirc$                                & $\times$                             & $\bigcirc$           \\ \hline
		Gravitational anomaly & $\bigcirc$                                & $\times$                              & $\bigcirc$           \\ \hline
		Extended monopoles and dyons & $\times$                                & $\bigcirc$                              & $\bigcirc$           \\ \hline
	\end{tabular}
	\renewcommand{\arraystretch}{1}
	\caption{In this Table we have summarized the main differences and similarities between 2D, 4D and 6D FQHEs discussed in the paper. The symbols $\bigcirc$ and $\times$ refer to the presence and absence of a specific feature, respectively.}
	\label{Table_one}
\end{table}

\section{Skyrmionic membranes}
\noindent Here, we are going to discuss the physics of the 6D FQHE in presence of a non-topological order parameter associated to a local symmetry breaking. This will allow us to show the existence of skyrmionic membranes in the bulk and discuss novel features of the boundary states.
In two spatial dimensions, the quantum Hall ferromagnetism can be understood from the field-theory point view through a coupling between a SO(3) vector associated to the ferromagnetic order and the hydrodynamic gauge fields \cite{Maciejko}. This picture can be naturally generalized to the six-dimensional case by replacing the SO(3) unit vector with a SO(5) unit vector $\phi_i$, with $i=1,2,3,4,5$ such that $\phi_i \phi_i =1$. As already shown in Refs\cite{Wu-Zee, Tze, Hasebe, Bernevig}, in 6+1 dimensions it is possible to define the following three-form current trough the SO(5) unit vector
\beq\label{Hall3}
J_{\rm Sk} =\frac{1}{8 \pi} \star {\rm \tr}\, \phi (d\phi)^4,
\eeq
where we have used a shorted notation $(d\phi)^4=  d\phi \wedge d\phi \wedge d\phi \wedge d\phi$ \cite{Hasebe}, with $\phi = \phi_i \Gamma_i$, where $\Gamma_i$ are the five Euclidean $4 \times 4$ Dirac matrices. 
This current describes skyrmionic membranes and generalizes the more common current for point-like skyrmions associated to SO(3). Moreover, the corresponding topological charge of these SO(5) skyrmions is given by $Q_{\rm Sk}= \mathcal{N} \int_{S^4} \star J_{\rm Sk}$ with $\mathcal{N}$ a suitable normalization factor.
The effective action for the SO(5) non-linear sigma coupled to the FQH phase is then given by
\begin{eqnarray}
S_{\rm {NL}}[c, C, \phi]=\int_{M_7} \left[-\frac{p}{4\pi} c \wedge dc +\frac{1}{2\pi} C \wedge dc+ J \wedge \star C + \right. \nonumber \\ \left. J_{\rm Sk} \wedge \star c +\frac{1}{2}{\rm \tr}\, d\phi \wedge \star d\phi   \right], \hspace{0.3cm}
\end{eqnarray}
where the last term is the (relativistic) kinematic term for $\phi$.
It is convenient at this point to employ the $HP^1$ formulation of the SO(5) non-linear sigma models \cite{Wu-Zee}. It is possible to represent $\phi_i$ in terms of a pair of quaternionic fields
$Q=(q_1, q_2)^T$ that satisfy the condition $Q^\dagger Q = 1$. We can then define an SU(2) one-form connection $\mathcal{A}= Q^\dagger d Q$ such that a novel three-form field is defined as
\beq
 \mathcal{C}={\rm \tr\, CS_3}(\mathcal{A}),
 \eeq
 with ${\rm CS_3}(\mathcal{A})= \mathcal{A} d \mathcal{A}+(2/3) \mathcal{A} \wedge \mathcal{A} \wedge \mathcal{A}$. In this way we obtain the following identity
\beq\label{Hall4}
 d \mathcal{C}=\frac{1}{4}\, {\rm \tr}\, \phi (d\phi)^4.
\eeq
Here, $\mathcal{C}$ can be seen as the order parameter gauge field. More in general, we can consider this model in a curved background by taking into account a topological coupling between $c$ and the geometric Chern-Simons terms introduced in the previous section. 
By integrating out the $c$ field, we then obtain the following new topological terms
\begin{eqnarray}\label{Hopf}
\int \mathcal{C} \wedge d \mathcal{C} + ({\rm CS}_3(\omega)+\beta {\rm CS}_3(e)) \wedge d\mathcal{C}.
\end{eqnarray}
The first term is nothing but the Hopf-Chern-Simons invariant, which is related to the second Hopf map $S^7 \rightarrow S^4$ and plays a central role in the characterization of the fractional statistics of membranes \cite{Wu-Zee, Tze}. The other terms are new and describe the coupling of the skyrmionic membranes to the curved background. 
The second Hopf invariant naturally generalizes the first Hopf map associated to point-like skyrmions in 2+1-D while in 4+1-D there are no Hopf maps. This fact, one more time, tells us that the 6D FQHE is the most natural generalization of the 2D FQHE. Similarly, the kinematic term of the non-linear sigma model can be completely rewritten in terms of the quanternionic fields and SU(2) one-form connection \cite{Wu-Zee}. 
The novel effective field theory in Eq. (\ref{Hopf}) should somehow give rise to novel features of the boundary. This should be the case in particular for open skyrmionic membranes for which their spatial boundaries live on the spatial boundary of the system and behave as skyrmionic strings.
To understand better this point, we embed the skyrmionic three-form field into a more general expression that takes into account also a local SO(5) gauge field $A$.
The corresponding new four-form field strength is then given by \cite{Intriligator, Radu, Nitta}
\beq\label{Hall5}
d \mathcal{C} \rightarrow d \mathcal{\hat{C}}={\rm \tr}\,  \phi \left[F + (1/2) (D_A \phi)^2\right]^2,
\eeq
where $F=d A+ A \wedge A$ and $D_A \phi = d\phi + A \wedge \phi$. This field strength recovers Eq. (\ref{Hall4}) for $A=0$ and represents the natural higher-form generalization of the field strength associated to SO(3) 't Hooft-Polyakov monopoles in 3D \cite{Hooft}.
We can directly express the novel three-form field $\mathcal{\hat{C}}$ as follows
\begin{eqnarray}\label{Hall6}
\mathcal{\hat{C}}={\rm \tr\, CS_3}(\mathcal{A})+ {\rm \tr}\, \phi \left[ (d \phi)^2 A + \right. \nonumber \\ \left.(1/2)(d \phi A \phi A + dA A+ AdA)+ \right. \nonumber \\
\left.(1/4)(A^3+ (1/3)A \phi A \phi A)\right],
\end{eqnarray}
where we remind that the first trace is taken on SU(2) while the second on SO(5) (here, $\phi$ and $\mathcal{A}$ are not independent fields).
This tensor field replaces $\mathcal{C}$ in Eq. (\ref{Hopf}) and as shown in Ref.\cite{Intriligator}, the corresponding Hopf-Chern-Simons term can be seen as a Wess-Zumino term of a suitable 5+1-D chiral superconformal field theory (SCFT). Its bosonic sector is characterized by a self-dual three-form field strength and a SO(5) scalar fields $\phi$. Thus, the Hopf-Wess-Zumino term takes into account the 't Hooft anomaly of CFT, which can exist only at the boundary of a 6+1-D system. 
As mentioned in the previous section and in Ref. \cite{Heckman}, the 6D FQHE naturally supports chiral (S)CFTs on its boundary that are characterized by chiral
two-forms, i.e. two-forms with self-dual or anti-selfdual field strength. We then conclude by conjecturing that the bosonic sector of the higgsed $N=(2,0)$ SCFT in Ref.\cite{Intriligator} represents a natural boundary theory of the 6D FQHE in presence of skyrmionic membranes.

\section{Dyonic strings}
\noindent In this section, we discuss the physical conditions that allows us to obtain dyonic strings in the bulk of the 6D FQH states, namely magnetically charged strings that acquire an electric charge through a generalized Witten effect.\\
For generic open membranes, $d \star J= J_\mathcal{B}$, where $J_\mathcal{B}$ is a two-form current. Here, $J_{\mathcal{B}}$ represents the physical current of the boundary the membranes, i.e. one-dimensional strings that live on the boundary of the system and couple with the two-form field $\mathcal{B}$.
In Ref. \cite{Intriligator}, it was argued that besides the generalized Hopf-Wess-Zumino term discussed in previous section, the effective action of the boundary theory also contains another topological term that governs the coupling of the skyrmionic strings to the two-form field $\mathcal{B}$, namely
\begin{equation}\label{matter}
S_{\rm {Sk-\mathcal{B}}}=\frac{\alpha}{2}\int_{\partial M_7} \mathcal{B} \wedge d \mathcal{\hat{C}} = \frac{\alpha}{2}\int_{M_7} d \mathcal{B} \wedge d \mathcal{\hat{C}},
\end{equation}
where $\alpha$ is a suitable constant. Thus, the skyrmionic current $\hat{J}_{\rm Sk}= (1/2 \pi) \star d \mathcal{\hat{C}}$ behaves as electric and magnetic source. This is in agreement with the fact that in 5+1-D, strings, in presence of self-dual three-form field strengths, become dyons \cite{Deser}.
The underlying mechanism is nothing but the higher-dimensional version of the Witten effect that occurs in 3+1-D \cite{Witten3, Franz}, where point-like magnetic monopoles acquire an electric charge through an axion field.
Thus, to obtain dyonic strings in the bulk of our 6D system, we need to first consider magnetic strings in our bulk and then introduce an axion field by generalizing the Witten effect to our odd spacetime dimensional case (the conventional Witten effect is defined in even spacetime dimensions).
In 6D, static magnetic strings can behave as extended magnetic monopoles \cite{Nepomechie, Teitelboim}. Thus, let us suppose that the static magnetic string has coordinate $(0,0,0,0,0, x^6)$, namely it is a line string spanned by $x^6$. Its corresponding four-form field strength is then given by
\begin{equation} \label{potential}
D_{ijkl}= \frac{3 Q_M}{2 \pi r^5} \epsilon_{ijklm 6} x^m,
\end{equation}
where $r=\sqrt{x_i x^i}$, with $i,j,k,l,m= 1,2,3,4,5$ and $Q_M$ is the magnetic charge. We have then that $\int_{S^4} D = 4 \pi Q_M$, where $S^4$ is the four-dimensional sphere with $r^2 =1$ that surrounds the extended magnetic monopole. We assume that there exists a suitable configuration of $\mathcal{\hat{D}}=d \mathcal{\hat{C}}$ that supports the above static gauge potential. This physically implies that suitable static three-form fields can be also related to magnetic string-like objects and not just to membranes. 
In order to show that these magnetic strings become dyons, we consider the following action in flat space
\begin{eqnarray}
S_{\rm {W}} [\theta, \mathcal{B}, \mathcal{\hat{C}}] = \int_{M_7}\frac{1}{4 \pi^2}\, \theta\, d \mathcal{B} \wedge d \mathcal{\hat{C}}+\frac{1}{ 4 \pi p}\mathcal{\hat{C}} \wedge d \mathcal{\hat{C}} \nonumber \\ -\frac{1}{4}\, d \mathcal{B} \wedge \star d \mathcal{B} + \mathcal{B}\wedge \star J_\mathcal{B}, \hspace{0.2cm}
\end{eqnarray}
where we have promoted $\mathcal{B}$ to be a dynamical gauge field that couples to a two-form current $J_\mathcal{B}$. Moreover, $\theta$ is an axion field. We have omitted for simplicity all the other terms which are not relevant for our current discussion.
From now on, we will threat $\mathcal{\hat{C}}$ as an independent gauge field instead than a composite gauge connection, such that it will formally behave as a dynamical background field like $C$.
By varying the action with respect to $\mathcal{B}$, we obtain
\begin{align}
d \star d \mathcal{B} +\frac{1}{4 \pi^2}\, d\theta\, \wedge d \mathcal{\hat{C}} = \star J_{\mathcal{B}}.
\end{align}
We now assume for simplicity that the axion field is only time-dependent $\theta=\theta(t)$ and homogeneous in space \cite{Franz} and that only the charge density is non-zero in $\star J_{\mathcal{B}}$, namely $J_{\mathcal{B}}^{\mu\nu}=(\rho^{i0}, 0)$, with $i=1,2,3,4,5,6$. The components of the above equations are given by 
\begin{align}
\partial_j \bar{E}^{ij}= \rho^{i}, \nonumber \\
-\partial_0 \bar{E}^{ij}+ \epsilon^{ijk lmn} \partial_k \bar{B}_{lmn}+\frac{1}{4 \pi^2} \partial_0 \theta \hat{B}^{ij}=0,
\end{align}
where $\hat{B}^{ij}=\epsilon^{ijklmn} \mathcal{\hat{D}}_{klmn}$, $\bar{E}_{ij} = \mathcal{H}_{ij0}$ and $\bar{B}^{ijk}=\epsilon^{ijklmn} \mathcal{H}_{lmn}$.
By taking the divergence of the second equation and combining it with the first one, we obtain
\begin{align}
\partial_0  \rho^{i} = \frac{1}{4 \pi^2} \partial_0 \theta \partial_j \hat{B}^{ij}.
\end{align}
We now insert the static gauge potential in Eq.(\ref{potential}) in the above equation such that the only non-zero component is given by
\begin{align}
\partial_0  \rho^{6} \equiv\partial_0  \rho = \frac{1}{4 \pi^2} \partial_0 \theta \partial_j \hat{B}^{6 j}\equiv
\frac{1}{4 \pi^2} \partial_0 \theta\, Q_M \delta^5(x),
\end{align}
where $\delta^5(x)$ is the Dirac delta function. We can now integrate on time and on the five-dimensional space to obtain the relation between electric and magnetic charges, namely
\begin{align}
Q_E = \frac{1}{2 \pi} \Delta\theta\, Q_M,
\end{align}
If we assume that there
was no initial electrical charge bound to the magnetic string monopoles then $\Delta \theta=\{0,\pi\}$.
This implies that
\begin{eqnarray}
Q_E= \frac{1}{2}  Q_M.
\end{eqnarray}
In other words, a string-like magnetic monopole associated to the three-form field acquires an electric charge related to the two-form field through an axion field. Importantly, this 6+1-D Witten effect can be naturally extended in 4+1-D as we will show in the Appendix A.\\

\section{Conclusions and outlook}
\noindent Summarizing, we have discussed the important role of extended objects in the 6D FQHE. Here, membranes and strings are related to three-form and two-form gauge fields, respectively. These gauge fields replace the more familiar one-form fields that appear in the effective field theory description of the more 2D FQHE.
We have shown that skyrmionic membranes associated to SO(5) naturally couple with the curved background through a generalized version of the Wen-Zee term and become important in a better identification of the 5+1-D chiral CFT that lives on the boundary of the system.
Moreover, we have also shown that besides the strings that live on the boundary, also magnetic strings in the bulk can transmute to dyonic strings, i.e. they can acquire an electric charge through a generalized Witten effect that involves an axion field and a dynamical two-form (Kalb-Ramond) field.
Although these novel results have been derived through effective quantum field theories in the infrared limit, their physical implications are general and could be also relevant in higher-dimensional microscopic systems where suitable lattice deformations/defects or generalized vortices \cite{Price, Pretko} in superfluid phases can behave as string-like and membrane-like objects.

\vspace{0.3cm}

\noindent {\bf Acknowledgments: }
The author is pleased to acknowledge discussions with Dimitra Karabali and Parameswaran Nair.

\section{Appendix A}
\noindent A 4D FQHE for strings cannot be properly built mainly because of the absence of a gravitational anomaly and because the Chern-Simons terms such as $\mathcal{B} \wedge d \mathcal{B}$ are total derivatives. Moreover, there is not a natural extension of the Wen-Zee term that couples the curved background and two-forms in a consistent way.
However, we can still build skyrmionic currents and study dyonic strings in a 4D FQHE that involves also one-form gauge potentials as we show in this appendix. A SO(4) order parameter allows us to build a two-form current for skyrmionic strings, given by
\beq\label{Hall7}
J_{\rm Sk} =\frac{1}{8 \pi} \star {\rm \tr}\, \phi (d\phi)^3,
\eeq
$\phi = \phi_i \Gamma_i$, where $\Gamma_i$ are four Euclidean $4 \times 4$ Dirac matrices. 
This current can be directly coupled to a two-form potential: $ \mathcal{B} \wedge \star J_{\rm Sk} $.
Due to the absence of an Hopf map, we cannot rewrite this current in terms of a two-form gauge field.
We can then introduce the following action in 4+1-D
\begin{eqnarray}
S [\theta, \mathcal{B}, \mathcal{A}] = \int_{M_{5}}\frac{1}{4 \pi^2}\, \theta\, d \mathcal{B} \wedge d \mathcal{A}+{\rm CS}_5(\mathcal{A}) \nonumber \\ -\frac{1}{4}\, d \mathcal{B} \wedge \star d \mathcal{B} + \mathcal{B}\wedge \star J_\mathcal{B}, \hspace{0.2cm}
\end{eqnarray}
where $J_\mathcal{B}$ is a generic two-form current and can also contain $J_{\rm Sk}$. Here, ${\rm CS}_5(\mathcal{A})$ is the 4+1-D Chern-Simons form built from a standard one-form field $\mathcal{A}$ (notice that its corresponding level is not important in our current discussion and for this reason it is encoded implicitly in the definition of ${\rm CS}_5(\mathcal{A})$).
 For simplicity we consider this action in flat space-time $M_5$ in order to avoid the geometric terms such as the Wen-Zee-like terms that can be constructed by coupling the curved background to one-form fields as shown in Ref.\cite{Karabali4}. Moreover, an axion field is crucial in order to show the existence of a generalized Witten effect. When $\theta$ is constant the first term in the above action becomes a total derivative and $\mathcal{A}$ and $\mathcal{B}$ are completely decoupled in the bulk. In this case we recover the standard 4D FQHE for point-like objects discussed in other works.
 By varying the above action with respect to $\mathcal{B}$, we obtain
 \begin{align}
 d \star d \mathcal{B} +\frac{1}{4 \pi^2}\, d\theta\, \wedge d \mathcal{A} = \star J_{\mathcal{B}}.
 \end{align}
 As done in 6+1-D case, we now assume for simplicity that the axion field is only time-dependent $\theta=\theta(t)$ and homogeneous in space and that only the charge density is non-zero in $\star J_{\mathcal{B}}$, namely $J_{\mathcal{B}}^{\mu\nu}=(\rho^{i0}, 0)$, with $i=1,2,3,4$. The components of the above equations are given by 
 \begin{align}
 \partial_j \bar{E}^{ij}= \rho^{i}, \nonumber \\
 -\partial_0 \bar{E}^{ij}+ \epsilon^{ijk l} \partial_k \bar{B}_{l}+\frac{1}{4 \pi^2} \partial_0 \theta \tilde{B}^{ij}=0,
 \end{align}
 where $\tilde{B}^{ij}=\epsilon^{ijkl} \mathcal{F}_{kl}$, $\bar{E}_{ij} = \mathcal{H}_{ij0}$ and $\bar{B}^{i}=\epsilon^{ijkl} \mathcal{H}_{jkl}$.
 By taking the divergence of the second equation and combining it with the first one, we obtain
 \begin{align}\label{Witten}
 \partial_0  \rho^{i} = \frac{1}{4 \pi^2} \partial_0 \theta \partial_j \tilde{B}^{ij}.
 \end{align}
In 4D, static magnetic strings can behave as extended magnetic monopoles \cite{Nepomechie}. Thus, let us consider that the static magnetic string has coordinate $(0,0,0, x^4)$, namely it is a line string spanned by $x^4$. Its corresponding two-form field strength is then given by
\begin{equation} \label{potential2}
F_{ij}= \frac{Q_M}{ r^3} \epsilon_{ijk x^4} x^k,
\end{equation}
where $r=\sqrt{x_i x^i}$, with $i,j,k= 1,2,3$ and $Q_M$ is the magnetic charge. We have then that $\int_{S^2} F = 4 \pi Q_M$, where $S^2$ is the two-dimensional sphere with $r^2 =1$ that surrounds the extended magnetic monopole. By inserting the above monopole field strength in Eq. (\ref{Witten}), we obtain
\begin{align}
\partial_0  \rho^{4} \equiv\partial_0  \rho = \frac{1}{4 \pi^2} \partial_0 \theta \partial_j \tilde{B}^{4 j}\equiv
\frac{1}{4 \pi^2} \partial_0 \theta\, Q_M \delta^3(x),
\end{align}
where $\delta^3(x)$ is the Dirac delta function. We can now integrate on time and on the three-dimensional space to obtain the relation between electric and magnetic charges, namely
\begin{align}
Q_E = \frac{1}{2 \pi} \Delta\theta\, Q_M,
\end{align}
If we assume that there
was no initial electrical charge bound to the magnetic string monopoles then $\Delta \theta=\{0,\pi\}$.
This implies that
\begin{eqnarray}
Q_E= \frac{1}{2}  Q_M.
\end{eqnarray}
In other words, string-like magnetic monopoles associated to a one-form field acquire electric charge related to the two-form field $\mathcal{B}$ through an axion field.\\

\appendix

\bibliography{ref}






\end{document}